%% file: ms_arp147.tex
\documentclass[iop]{emulateapj}

\usepackage{graphicx}
\usepackage{color}

\newcommand{\err}[2]{\ensuremath{^{_{+#1}}_{^{-#2}}}}
\newcommand{\ee}[2]{\ensuremath{{#1}\!\times\!10^{#2}}}

\newcommand{\ergcmsa}{\ensuremath{\mathrm{erg~cm}^{-2}\,\mathrm{s}^{-1}\mathrm{\AA}^{-1}}}

\newcommand{\pcmsq}{\ensuremath{\mathrm{cm}^{-2}}}
\newcommand{\cps}{\ensuremath{\mathrm{counts~s}^{-1}}}

\newcommand{\galex}{\textit{Galex}}

\begin{document}

\shortauthors{Rappaport, Levine, Pooley, \& Steinhorn}
\shorttitle{Arp 147 Colliding Galaxies}
\title{Luminous X-Ray Sources in Arp 147}

\author{S. Rappaport\altaffilmark{1}, A. Levine\altaffilmark{2},
D. Pooley\altaffilmark{3}, B. Steinhorn\altaffilmark{1}}

\altaffiltext{1}{37-602B, M.I.T. Department of Physics and Kavli
 Institute for Astrophysics and Space Research, 70 Vassar St.,
 Cambridge, MA, 02139; sar@mit.edu} 
 \altaffiltext{2}{37-575 M.I.T. Kavli
 Institute for Astrophysics and Space Research, 70 Vassar St.,
 Cambridge, MA, 02139; aml@space.mit.edu} 
 \altaffiltext{3}{Eureka Scientific, 5248 Valley View Road
El Sobrante, CA 94803-3435; pooley@gmail.com}

\begin{abstract}
The {\em Chandra X-Ray Observatory} was used to image the collisional
ring galaxy Arp 147 for 42 ks.  We detect 9 X-ray sources with
luminosities in the range of $1.4 - 7 \times 10^{39}$ ergs s$^{-1}$
in or near the blue knots of star formation associated
with the ring.  A source with an isotropic X-ray luminosity of $1.4 \times 10^{40}$ ergs s$^{-1}$
is detected in the nuclear region of the intruder galaxy.  
X-ray sources associated with a
foreground star and a background quasar 
are used to improve the registration of the X-ray
image with respect to {\em HST} high resolution optical images.  The intruder 
galaxy, which apparently contained little
gas before the collision, shows no X-ray sources other than
the one in the nuclear bulge which may be a poorly fed supermassive
black hole.  These observations confirm the conventional wisdom that
collisions of gas rich galaxies trigger large rates of star formation
which, in turn, generate substantial numbers of X-ray sources, some of
which have luminosities above the Eddington limit for accreting stellar-mass
black holes.
\end{abstract}


\section{Introduction}

\begin{figure*}[h!]
\vspace{0.3cm}
\begin{center}
\includegraphics[width=0.9 \textwidth]{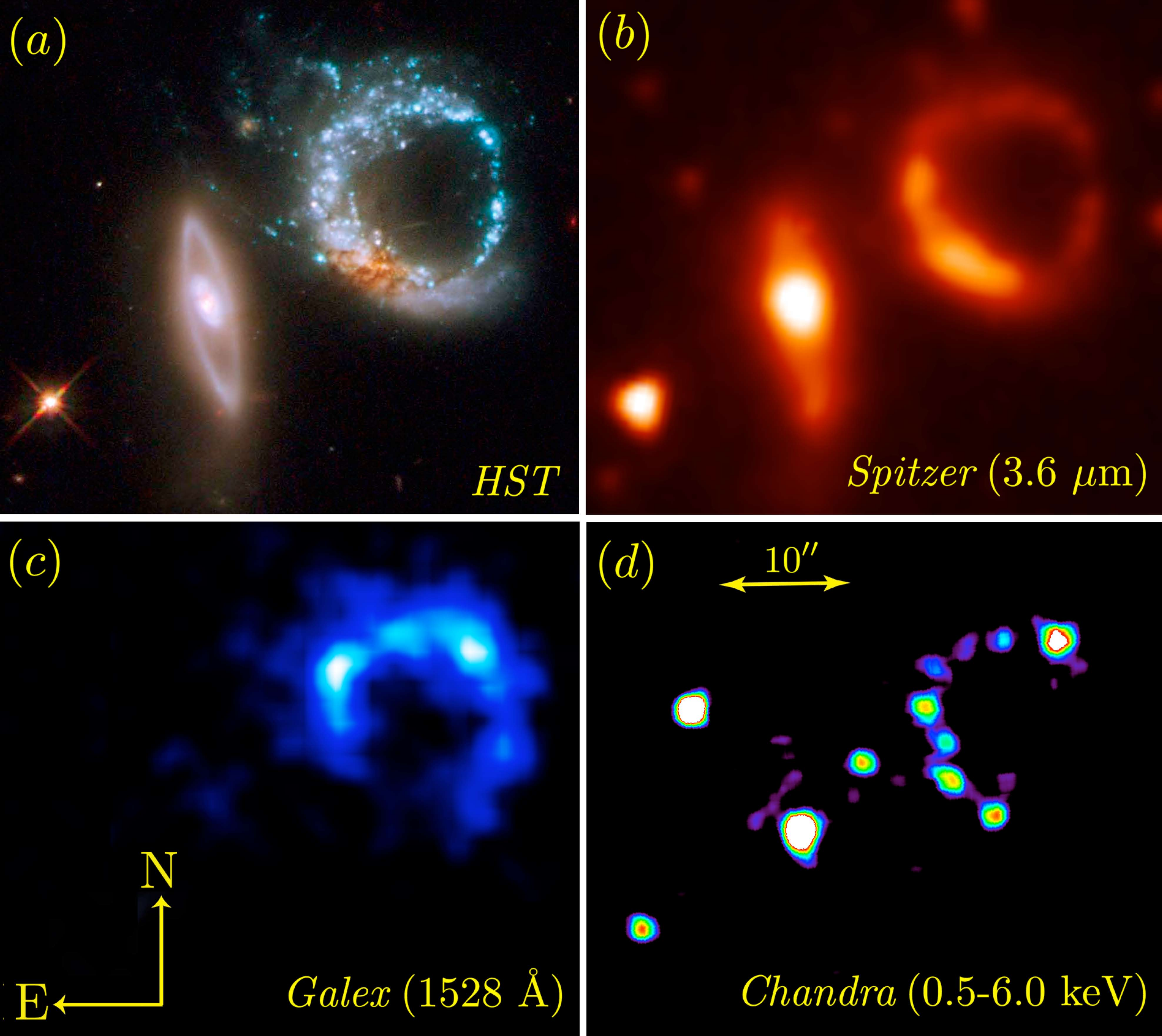}
\caption{Montage of images taken of the collisional ring galaxy Arp 147.  {\em Panel (a)}: {\em HST} image taken with WFPC2 (at 4556 \AA, 5997 \AA, and 8012 \AA). Note the intense blue ring associated with star formation as well as the reddish intruder galaxy to the southeast (also exhibiting a ring structure). The blue ring is $\sim$17$''$ in diameter, which corresponds to $\sim$11 kpc at the distance of 133 Mpc.   {\em Panel (b)}: the corresponding {\em Spitzer} image of Arp 147 in the 3.6 $\mu$m band.  {\em Panel (c)}: {\em Galex} 1516 \AA \,image from archival data.  {\em Panel (d)}: 42 ks {\em Chandra} image (this work).  Aside from 8 discrete luminous X-ray sources associated with the ring of star formation, there are 4 other point sources.  From left to right, these correspond to a foreground star, a distant galaxy, the nucleus of the intruder galaxy, and a blue knot that happens to be offset from the main ring.}
\label{fig:Arp147}
\end{center}
\end{figure*}

If one or both of a pair of colliding galaxies has a high gas content,
the collision may trigger a spectacular burst of star formation like
those seen in the Antennae (Whitmore \& Schweizer 1995) and the
Cartwheel (e.g., Higdon 1995; Amram et al.~1998).  In the latter
case, a smaller intruder galaxy passed through the disk of
the progenitor spiral galaxy about $3 \times 10^8$ years ago.  This
triggered a wave of star formation which has propagated radially
outward at an effective speed of $\sim$60 km s$^{-1}$ and is
apparent as a brilliant, expanding ring.  The dynamics of this
so-called collisional ring galaxy and other colliding galaxies,
including examples that are remarkably similar in appearance to the
Cartwheel, have been explored and understood at least in part through
numerical simulations (see, e.g., Lynds \& Toomre 1976; Toomre 1978;
Gerber, Lamb, \& Balsara 1992; Mihos \& Hernquist 1994; Appleton \&
Struck 1996).

Many of the newly formed stars in a collision-induced ring or other
star formation region must naturally constitute binary systems.  Some fraction
of the more massive stars in these binaries evolve rapidly to form
conventional high-mass X-ray binaries (with neutron-star or
stellar-mass black-hole accretors and $L_x \lesssim 10^{39}$ ergs
s$^{-1}$) that can be detected with {\em Chandra} out to distances
of $\sim$50 Mpc (see, e.g., Fabbiano 2006).  If the examples of the
Antennae (Zezas \& Fabbiano 2002) and the Cartwheel (Gao et al.~2003;
Wolter, Trinchieri, \& Colpi 2006) are representative, galaxy
collisions also produce substantial numbers of ultraluminous X-ray
sources (``ULXs'') which are characterized by $L_x \gtrsim 2 \times
10^{39}$ ergs s$^{-1}$. The nature of the ULXs is unclear at
present; they are likely to be binaries with black-hole accretors, but
the accretors could be either of stellar mass and accreting at rates
far above the Eddington limit or of masses one to two orders of
magnitude higher and be so-called intermediate-mass black holes
(``IMBHs''; see, e.g., Colbert \& Mushotzky 1999).

At the higher end of the ULX luminosity function, i.e., at $L_x
\gtrsim 10^{40}$ ergs s$^{-1}$, and especially as the inferred
luminosities approach $\sim$$10^{41}$ ergs s$^{-1}$, it becomes
increasingly difficult to see how the requisite emission, even if
somewhat beamed, could be produced around a stellar-mass black hole
(see, e.g., Madhusudhan et al.~2008).  By contrast, the X-ray luminosities of
accreting IMBHs could easily exceed $10^{41}$ ergs s$^{-1}$ without
violating the Eddington limit, but this explanation of ULXs is
confronted by other serious problems.  Portegies Zwart et al.~(2004),
among others, have proposed that runaway star collisions in newly
formed massive star clusters lead to the formation of supermassive
stars (e.g., $\gtrsim 500~M_\odot$) which, in turn, evolve to form
IMBHs.  The IMBHs must then capture massive stars into orbits where
mass transfer will proceed at levels sufficient to produce the
requisite X-ray emission.  However, the
evolution of supermassive stars is highly uncertain, and the
efficiency for producing the requisite numbers of IMBHs in the
Cartwheel, for example, is implausibly high (King 2004).  A better
understanding of ULXs could shed much light on these issues and on the
formation and evolution of very massive stars.
	
Since it appears that substantial numbers of ULXs are commonly found
in collisional ring galaxies, it would seem like these would be good
targets for {\em Chandra} observations.  The ``Atlas and Catalog of
Collisional Ring Galaxies'' (Madore, Nelson, \& Petrillo 2009)
contains information on $\sim$$104$ collisional ring galaxies.  While
this atlas focuses on southern hemisphere objects, it also includes
information on northern hemisphere collisional ring galaxies that are
described in the literature including the Arp peculiar galaxy catalogs
(Arp 1966; Arp \& Madore 1987).  We find that only 4 of the $\sim$$104$
collisional ring galaxies have been observed by {\em Chandra}.  One of
these is the Cartwheel galaxy. The others are Arp 284, Arp 318, and AM
0644$-$741, but Arp 318 was observed at an off-axis angle of
$>4\arcmin$ so the existing data would not be useful in resolving
individual point sources.  Results from the observation of Arp 284
have been published by Smith et al. (2009) who report the detection of
$\sim$7 ULXs, with two having $L_x \sim 10^{40}$ ergs s$^{-1}$.

In July 2009, we obtained {\em Chandra} observations of the well-known
collisional ring galaxy Arp 147.  An image of Arp 147 taken earlier with the
WFPC2 on {\em HST} is shown in Fig.\,1a.  The object consists of the
ring-like remnant of what was originally a gas-rich galaxy that likely
underwent, according to Gerber et al.\,(1992), an off-center collision
with an approximately equal-mass ($\sim$$1.8 \times 
10^{11}\,M_\odot$) elliptical galaxy  that passed ``perpendicular
through the disk about two radial scale lengths from the
center''. However, simulations of similar systems suggest that the
mass of the intruder could be $2-3$ times more massive than the ring
(A. Toomre, private communication).  Hereafter, we refer to the
ring-like structure exhibiting a high star formation rate (see, e.g.,
the blue knots in Fig.~1a) as the ``ring'', and to the reddish
neighbor galaxy, which also appears to have a tidally induced ring
structure, as the ``intruder''.  The maximum angular extents of the
ring and the intruder galaxies are $\sim$$18-19\arcsec$, and their 
recession velocities are 9415 and 9656 km
s$^{-1}$, respectively.  For an assumed Hubble constant of $H_0 = 72$
km s$^{-1}$ Mpc$^{-1}$, the distance to Arp 147 is $\sim$133 Mpc, and
the corresponding physical size of the two objects is $\sim$12 kpc.  
Finally, we note that the reddish-pink region seen in the
{\em HST} image on the south-southeast side of the ring is thought to be
the nucleus of the original spiral galaxy which the intruder has
tidally stretched into an apparently ring-like structure (e.g., Gerber
et al.~1992).

In \S2.1 of this work we describe the {\em Chandra} X-ray observations
and the analysis of the data.  The supplemental images from the {\em
HST}, {\em Spitzer}, and {\em Galex} observatories are described in
\S\S2.2-2.4.  In \S3 we present the results of the X-ray observations
including a table that lists source locations, fluxes, luminosities, median 
photon energies, and power-law spectral indices.  The star formation 
rate (``SFR'') in the ring
portion of Arp 147 is inferred from, and contrasted among, the 
{\em Spitzer} and {\em Galex} fluxes as well the number of luminous 
X-ray sources.

\section{Observations}

\subsection{Chandra Observations}
\label{sec:xrayobs}

Arp 147 was observed with {\em Chandra} with the Advanced CCD Imaging
Spectrometer (ACIS) at the focal plane.  The data were taken in
timed-exposure mode with an integration time of 3.24 s per frame, and
the telescope aim point was on the backside-illuminated S3 
chip.  A 24.5 ks exposure (ObsID 11280) was made on 2009 September 13 and an 18.0 ks
exposure (ObsID 11887) was made on 2009 September 15 for a total exposure of 42.5
ks.  

The data from the two observations were merged and were reprocessed without the pixel randomization that is included during standard processing.  Reprocessing the data in this way slightly improves the point spread function.  An X-ray image was then formed from the events in the energy band
0.5-6.0 keV.  The image was smoothed for aesthetic purposes via a convolution with a two-dimensional Gaussian function (with $\sigma = 1.1\arcsec$).  The resultant X-ray image can be seen in panel (d) of Fig.~\ref{fig:Arp147}. 

Source detection was performed both with the wavdetect tool provided with the CIAO software\footnote{\url{http://asc.harvard.edu/ciao/}} and by visual inspection of a Gaussian-smoothed image.  A total of 9 statistically significant candidate sources around the ring were found, in addition to a nuclear source in the intruder galaxy, a background quasar, and a foreground star in the vicinity of Arp 147.  Light curves and spectra, as well as spectral response files (ARFs and RMFs) and background spectra, were extracted for each candidate source with the ACIS Extract package (AE; Broos et al.\ 2010).  The light curves were examined for variability, but nothing significant was found.  Source positions were refined in AE by using the mean photon position as the best estimate of the source's true location.  The uncertainty in each source's net counts was calculated using the method of Kraft, Burrows, \& Nousek (1991). Source positions and net counts are listed in Table~\ref{tab:xray}.

The source and background spectra were analyzed with Sherpa 4.2 (Freeman, Doe, \& Siemiginowska 2001).  AE was used to combine each source's spectrum and response files from both ObsIDs. In all of the analyses, the unbinned spectra in the 0.5--8 keV range were fit using the CSTAT statistic, a slight modification of the Cash (1979) statistic, and using the Nelder-Mead optimization algorithm (Nelder \& Mead 1965).  First, the background spectrum for each source was fit with an absorbed power law model.  Then, the source spectrum was fit with an absorbed power law model plus a background component based on the background spectrum fit.  In all cases, the source spectrum had so few counts that the column density was frozen at the Galactic value in the direction of Arp 147 of \ee{6.2}{20}~\pcmsq.  The unabsorbed fluxes and luminosities are reported in Table~\ref{tab:xray}.  The confidence intervals on the fluxes were found by sampling the distribution of power-law parameters $10^4$ times and calculating the unabsorbed flux from each set of sampled parameters (the samples of the power-law index were restricted to the range 0--6).  The median and standard deviation of the resultant flux distribution are given as the source's flux and uncertainty in Table~\ref{tab:xray}.  The power-law index is also reported as a measure of each source's spectral shape. Finally, we also list in Table~\ref{tab:xray} the median photon energy which, while somewhat instrument-dependent, yields a more statistically significant intercomparison of spectral hardness among the sources.

\subsection{Hubble Space Telescope Observations}

HST data obtained from an observation (ID 11902) of Arp 147 made on
2008 October 29 were retrieved from the STScI MAST \footnote{\url{http://archive.stsci.edu/}} data archive.  High level science products were provided by the Hubble Heritage Team as part of the Merging and Interacting Galaxies project\footnote{\url{http://archive.stsci.edu/prepds/merggal/}}.   The
WFPC2 instrument was used for the observation, and exposures were made
with the F450W (4556 \AA), F606W (5997 \AA), and F814W (8012 \AA) filters
for 8800, 6600, and 6600 seconds respectively.  The drizzled images had
intensity values in counts s$^{-1}$.  We used a large extraction region completely enclosing the Arp 147 ring to determine net count rates of 271, 1095, and 575 counts s$^{-1}$ in the F450W, F606W, and F814W filters, respectively.  Similarly, the intruder galaxy had net count rates of 204, 1230, and 917 counts s$^{-1}$.  These count rates were converted to fluxes using the PHOTFLAM conversion factor found in the header of each FITS image, which was \ee{9.02}{-18} for the F450W image, \ee{1.90}{-18} for the F606W image, and \ee{2.51}{-18} for the F814W image.

\begin{figure*}
\begin{center}
\vglue1cm
\includegraphics[width=0.68 \textwidth]{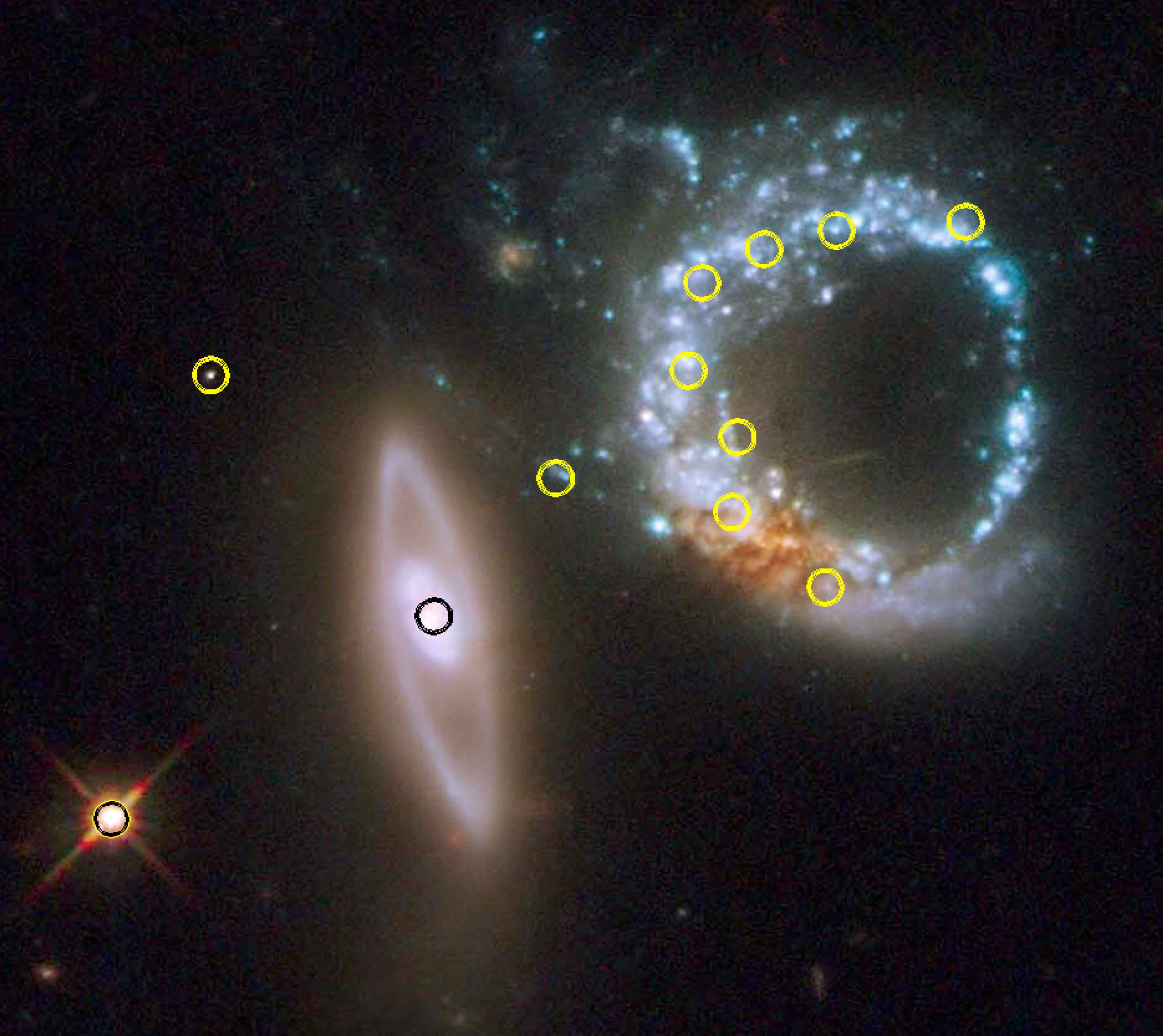}
\caption{HST image of the Arp 147 region with the locations of 12 detected X-ray sources marked with circles.  The foreground star to the lower left in the image, and the more distant (point-like) galaxy above it, were used to tie the Chandra locations to the HST image.  We estimate that the resultant astrometric alignment is probably good to between 1/3$''$ and 1/2$''$, or roughly half the radius of the orange circles.  The alignment between the two images is clearly good enough to show a strong correlation between the blue regions of star formation and the X-ray sources, but insufficiently precise to pinpoint an association between a given X-ray source and a particular blue knot.}
\label{fig:HSTXray}
\end{center}
\end{figure*}

\input{Arp147.tbl}

\subsection{Spitzer Observations}

A Spitzer Space Telescope observation (ID 20369) of Arp 147 was
performed on 2005 August 8 with the IRAC instrument.  We retrieved from the Spitzer archive the 3.6, 4.5, 5.8, and 8.0 $\mu$m post Basic Calibrated Data mosaic images using the Spitzer Leopard 
software\footnote{\url{http://ssc.spitzer.caltech.edu/warmmission/propkit/spot/}}.
The total exposure time for each wavelength band was 643 seconds (24 exposures, each of 26.8 s). We also retrieved the 24.0 $\mu$m image acquired with MIPS instrument on 2006 February 20, which had a total exposure time of 420 seconds (42 exposures, each 9.96 s).  Both the IRAC and MIPS FITS files had
pixel values in units of MJy sr$^{-1}$.  The 3.6$\mu$m IRAC image is shown in
panel (b) of Fig.~\ref{fig:Arp147}.  We converted the images to units of MJy pixel$^{-1}$ based on the pixel sizes of 0\farcs6012 for the IRAC images and 2\farcs45 for the MIPS image.  We found net flux densities for the Arp 147 ring of 3.8, 2.8, 8.7, 27.5, and 58.8 mJy for the 3.6, 4.5, 5.8, 8.0, and 24 $\mu$m images, respectively.  Likewise, we found net flux densities for the intruder of  5.4, 3.5, 3.6, 5.3, and 7.3 mJy. 

These {\em Spitzer} spectral points are summarized in Fig.\,\ref{fig:flam} as spectral luminosities.

\subsection{\galex\ Observations} 

We used GalexView 1.4\footnote{http://galex.stsci.edu/galexview} to retrieve data from the \galex\ archive.  Tile 181 of the All-Sky Imaging Survey (AIS 181) contains image data on Arp 147 in the near-UV (1771--2831\,\AA) and far-UV (1344--1786\,\AA) bands.  These bands have effective bandwidths of 732\,\AA\ (NUV) and 268\,\AA\ (FUV) and effective wavelengths of 2267\,\AA\ and 1516\,\AA. The observation occurred on 2007 October 20 and had an exposure of 216 s in each band.  We used the count maps in both bands and defined a 30\arcsec\ radius circular region to extract the counts from both the Arp 147 ring and the intruder galaxy, which are not completely separated by the \galex\ angular resolution.  We used a source-free annulus with inner radius 45\arcsec\ and outer radius 120\arcsec\ for our background estimate.  

In the NUV image, the net counts from the 30\arcsec\ inner region are 2603 $\pm$ 58, which we attribute to both the ring and the intruder.  Using two smaller, non-overlapping elliptical regions (of $\sim17'' \times 21''$ and $6'' \times 13''$ in size), we extracted $2247 \pm 50$ net counts from the ring and $193\pm16$ net counts from the intruder, respectively.  This left 163 net counts unaccounted for.  We divided those up proportionally, for a final result of $2397\pm 60$ net counts for the ring and $206\pm20$ net counts for the intruder.  To estimate the count rate, we took the average value of the NUV relative response high resolution (``rrhr'') map in the vicinity of Arp 147, which was 150 s.  The rrhr map combines the relative sensitivity of the detector with the exposure time.  Using the conversion factor from count rate to flux density in Table 1.1 of the \galex\ Observer's Guide of $\ee{2.05}{-16}~(\ergcmsa)/(\cps)$, we calculate a flux density of $(3.3\pm0.08) \times 10^{-15}~\ergcmsa$  for the ring and $(2.9\pm0.21) \times 10^{-16}$~\ergcmsa \, for the intruder.

For the FUV image, we follow the same procedure.  The net counts from the 30$''$ inner region are 795 $\pm$ 31, which we attribute to both the ring and the intruder.  Using two smaller, non-overlapping elliptical regions, we extracted $722 \pm 28$ net counts from the ring and $36.5\pm7.0$ net counts from the intruder.  This left 37 net counts unaccounted for.  We divided those up proportionally, for a final result of $757\pm29$ net counts for the ring and $38\pm9$ net counts for the intruder.  To estimate the count rate, we took the average value of the FUV `rrhr' map in the vicinity of Arp 147, which was 181.4 s.  Using the conversion factor from count rate to flux density in Table 1.1 of the \galex\ Observer's Guide of $\ee{1.40}{-15}~(\ergcmsa)/(\cps)$, we calculate a flux density of $(5.8 \pm0.2) \times 10^{-15}~\ergcmsa$  for the ring and \ee{(3 \pm0.7)}{-16}~\ergcmsa \, for the intruder.

\begin{figure}
\begin{center}
\vglue1cm
\includegraphics[width=0.49 \textwidth]{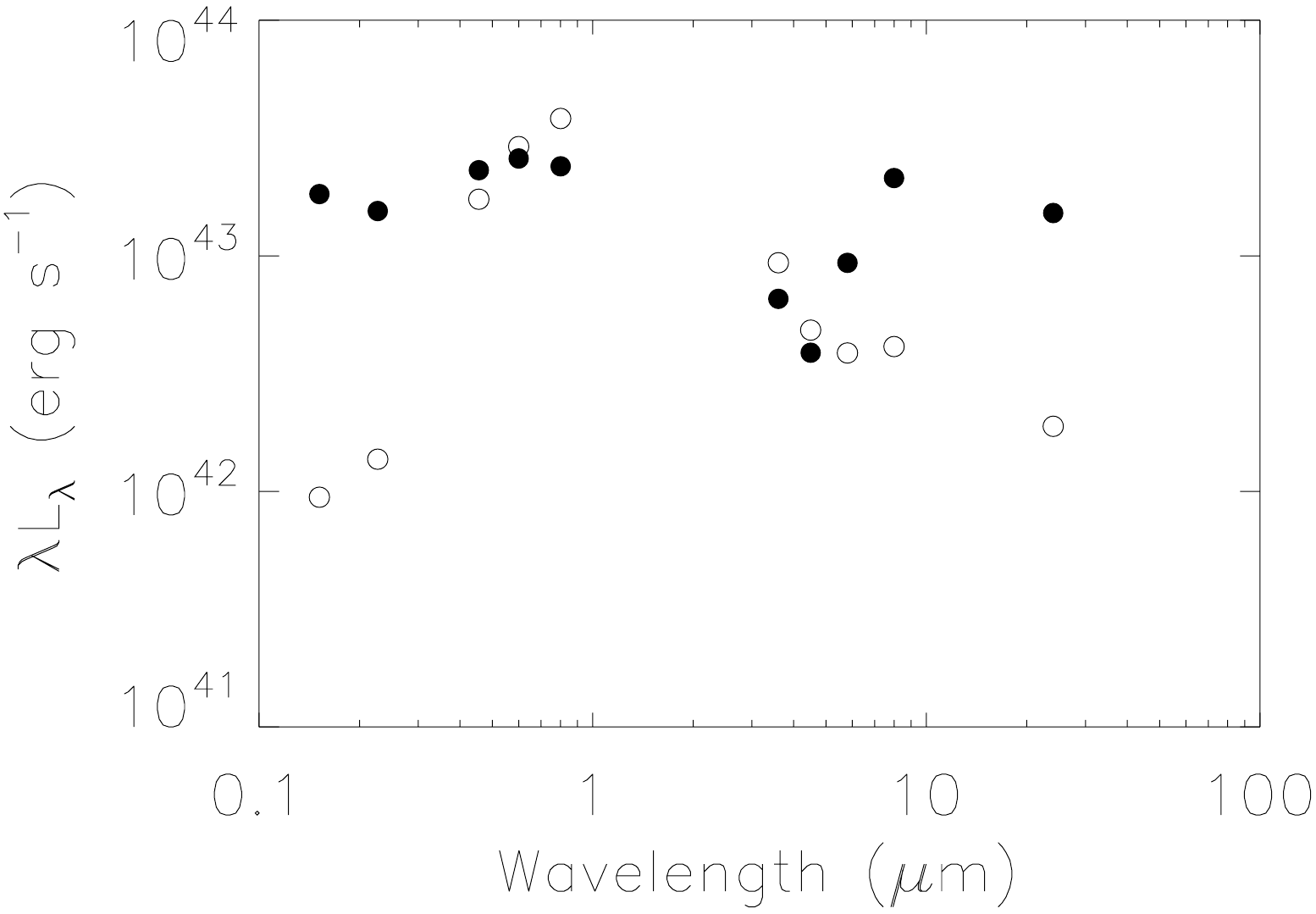}
\caption{Spectral luminosity distributions ($\lambda L_\lambda$) of the integrated light from the entire blue ring of Arp 147 (solid circles) and the entire intruder galaxy (open circles).  The points in the UV are from the {\em Galex} satellite, in the optical from {\em HST}, and in the IR from {\em Spitzer}.  The $\lambda L_\lambda$ values for the Arp 147 ring crudely indicate that the spectrum is approximately flat over 2.5 decades in wavelength, and that the total luminosity of the ring (in the $0.1-24\,\mu{\rm m}$ band) is $\sim$$2 \times 10^{43}$ ergs s$^{-1}$.}
\label{fig:flam}
\vspace{0.3cm}
\end{center}
\end{figure}

The spectral luminosities from both the ring and the intruder galaxy in the NUV and the FUV are shown in Fig.\,\ref{fig:flam}.

The FUV image, which has been smoothed via
convolution with a 2-dimensional Gaussian function (with $\sigma =
0.7\arcsec$) for aesthetic purposes, can be seen in panel (c) of
Fig.~\ref{fig:Arp147}.

\section{Results from the X-ray Observations}

In the {\em Chandra} image shown in Fig.~\ref{fig:Arp147}d, we find
eight discrete sources at locations around the blue ring plus a source
just outside the ring that falls close to or in a blue knot that must
be associated with the ring.  A
bright source is found within $\sim$$1/2\arcsec$ of the nucleus of the intruder
galaxy and two other sources are present at the locations of a bright foreground star and
a faint background object.  The latter two sources were used to align
the X-ray and optical images to an accuracy of $\sim$$1/2\arcsec$.  The
properties of the X-ray sources are summarized in Table 1 (see \S\ref{sec:xrayobs} for details). 
The locations of the X-ray sources are shown superposed on
the {\em HST} image in Fig.~2.

The faint background source is a 22nd magnitude object identified in
the {\em SDSS} archive as a galaxy (SDSS J031120.03+011858.4),
but which is very likely a quasar based on its X-ray luminosity.

The foreground star (SDSS J031120.42+011841.4) has $g = 18.3$, $u-g = 2.55$
and $g-r = 1.53$, and is likely a nearby M star (see, e.g., Smol\v{c}i\'c et al.\,2004).

We have carried out an autocorrelation analysis of the {\em HST} image
of the ring and find that the characteristic size of the blue knots is
$\sim$$0.6\arcsec$, i.e., a few hundred parsecs in physical size.  This
is also approximately the accuracy of our alignment of the X-ray image
to the {\em HST} image. While it is quite obvious that there are a
substantial number of ULX-level X-ray sources associated with the ring
of star formation, the $\sim$$1\arcsec$ resolution of the {\em Chandra}
image, the finite sizes of the blue knots in the {\em HST} image,
their abundance, and the finite precision of the X-ray to optical
image alignment prevent us from clearly identifying specific blue
knots with particular X-ray sources.  Therefore, we cannot positively
determine whether
the X-ray sources lie in or outside of individual knots.

Another issue is whether, in fact, each apparent source represents the
emission from a single object or a blend of fainter sources.  To
address this, we have carried out a Monte Carlo simulation using a
generic X-ray source luminosity function taken from Grimm, Gilfanov \&
Sunyaev (2003; $dN/dL_x \propto L_x^{-1.7}$). We choose the X-ray
source locations according to a uniform probability distribution
within an annular region that approximately represents the ring of
star formation.  Each simulation is designed to produce an image
containing the total X-ray flux that is actually observed.  The
simulated images are then smoothed with the {\em Chandra} point spread
function, and examined for detectable sources that may be, in fact,
composite.  Typically no more than 1 in 8 of our detected sources
within the ring can be expected to be comprised of multiple discrete
sources.  We have not, however, studied cases where the locations of X-ray
sources are highly correlated.

\section{The Star Formation Rate}
\label{sec:SFR}

There are a number of empirical relations between spectral fluxes in
different bands and star formation rates (``SFRs'').   
Given the excellent wavelength
coverage of Arp 147, we attempt to utilize those
relations to infer the star formation rate in the ring, and to compare
the results from different wavelengths.

A spectral energy distribution, integrated over the entire ring, is shown in
Fig.\,\ref{fig:flam}.  Here we have defined $L_\lambda \equiv 4 \pi d^2 F_\lambda$, where $F_\lambda$ is the spectral flux and $d$ is the distance to the source.  The quantity $\lambda L_\lambda$ has the dimensions of luminosity (and is equal to $\nu L_\nu$).  The two UV points are from the {\em Galex}
data, the three visible-region spectral points are from {\em HST}, and
the remaining five IR points are from {\em Spitzer}.

Infrared spectral intensities at
24 $\mu$m and at 8 $\mu$m can each be used to estimate the SFR (see, e.g., Wu et al.\,2005).  Equation
(1) of Wu et al.\,(2005) can be written as:
\begin{equation}
{\rm SFR}(24 \mu{\rm m}) \simeq \frac{\lambda \,L_\lambda}{6.7 \times 10^8\,L_\odot} ~~~M_\odot\,{\rm yr}^{-1}
\end{equation}
The 24 $\mu$m flux from the entire Arp 147 ring (see
Fig.\,\ref{fig:flam}) yields a SFR of $\sim$6 $M_\odot$ yr$^{-1}$.
Equation (2) of Wu et al.\,(2005) relates the flux in the 8 $\mu$m band to the SFR:
\begin{equation}
{\rm SFR}(8 \mu{\rm m}) \simeq \frac{\lambda \,L_\lambda}{1.4 \times 10^9\,L_\odot} ~~~M_\odot\,{\rm yr}^{-1}
\end{equation}
Thus, the 8 $\mu$m flux from the entire Arp 147 ring (see Fig.\,\ref{fig:flam}) yields a SFR of $\sim$4 $M_\odot$ yr$^{-1}$

The SFR can be estimated from the UV flux by using equation (12) of
Rosa-Gonzalez, Terlevich, \& Terlevich (2002):
\begin{equation}
{\rm SFR}(2000 \, {\rm \AA}) \simeq 6.4 \times 10^{-28} \,\frac{\lambda L_\lambda}{\nu \, {\rm (Hz)}} ~~ M_\odot {\rm yr}^{-1}.
\end{equation}
The approximate SFR from the {\em GALEX} measurement of the near UV for the Arp 147 ring is $\sim$6 $M_\odot$ yr$^{-1}$, in very reasonable agreement with the values obtained from the {\em Spitzer} measurements.

Finally, we can use the relationship between the number of luminous X-ray sources (i.e., with $L_x \gtrsim 2 \times 10^{38}$ ergs s$^{-1}$) and SFR given by Grimm, Gilfanov, \& Sunyaev (2003), in particular, their equation (20), to estimate the SFR:
\begin{equation}
{\rm SFR} \simeq 0.34 ~N(L>2 \times 10^{38} {\rm ergs~s}^{-1})
\end{equation}
The approximate SFR as inferred from the number of luminous {\em Chandra} X-ray sources in Arp 147 (with $L_x \gtrsim 1.4 \times 10^{39}$ ergs s$^{-1}$) is $\sim$12$\,M_\odot$ yr$^{-1}$.  This takes into account the fact that the current observations were only sensitive to sources a factor of about 7 more luminous than the Grimm et al.\,(2003) threshold luminosity, and an assumed differential luminosity function $\propto L^{-1.7}$ (as discussed in \S 3).

In any case, all the indicators point to a vigorous 
star formation rate within the ring of Arp 147 consistent with $\sim$6 $M_\odot$ yr$^{-1}$.  By comparison, the SFR of the Milky Way, which is presumably substantially more massive than the ring ($\sim$$8 \times 10^{11}\,M_\odot$ [to within a factor of $\sim$1.5] vs. $\sim$$1.8 \times 10^{11}\,M_\odot$; Battaglia et al.\,2006; Gnedin et al.\,2010), is $\sim$$4\pm 1\,M_\odot$ yr$^{-1}$ (see e.g., McKee \& Williams 1997; Diehl et al.\,2007, and references therein).

\section{Discussion and Summary}

Arp 147 is a visually impressive pair of galaxies that show the
effects of a recent collision.  The collision drew the less massive,
more gas-rich galaxy into a shape that appears ring-like, at least in
projection, and triggered large rates of star formation.  A dust-rich
region on the south-southeast side of the ring is likely to have
evolved from the nuclear region of the original gas-rich galaxy.  The
intruder galaxy is much redder in color and has a distinct ring
structure surrounding a bright nuclear bulge.  It seems reasonable to
assume that this ring was also induced by the collision.  Both the
color of the intruder galaxy and the absence of indicators of star formation in
its ring suggest that this galaxy was gas poor at the time of the
collision.

There are significant uncertainties regarding the geometry and time
scales of the collision even though there have been attempts, such as
those of Gerber et al.~(1992), to model the event through numerical
simulations.  Nonetheless, the observed radial velocities and
projected separation of the two galaxies suggest the scales of the
important parameters.  In particular, the recession velocities of the
two galaxies differ by $\sim$250 km s$^{-1}$, a difference that may be
taken to be loosely representative of the relative velocity projected
perpendicular to the line of sight.  Since the nucleus of the intruder
is separated (in projection) by $\sim$10 kpc from the original nucleus
of the ring, a relative speed of $\sim$250 km s$^{-1}$ would then
imply that closest approach occurred $\sim$40 Myr ago. 

For a Gaussian
speed distribution, it is straightforward to show that, given a radial speed,
$v_r$, the probability distribution for the tangential (i.e., sky plane) component,
$v_t$ is
\begin{equation}
p(v_t) \propto v_t \, \sqrt{v_t^2 + v_r^2}  \, \exp[-(v_t^2 + v_r^2)/2 \sigma^2]
\end{equation}
where $v_r = 250$ km s$^{-1}$ and we adopt a value of $\sigma = 235$ km s$^{-1}$
(see, e.g., Peebles 1980) since Arp 147 appears not to be part of any significant galaxy group.\footnote{
A $13' \times 13'$ box centered on the SDSS image of Arp 147 (corresponding
to a 1 Mpc square) contains only two galaxies that are fainter but of comparable 
brightness.} The most probable (relative) tangential velocity of Arp 147 based 
on the above probability distribution is $\sim$300 km s$^{-1}$, and the 80\% 
confidence upper and lower limits on $v_t$ are $\sim$475 km s$^{-1}$ and $\sim$200 km s$^{-1}$, respectively. 

At the former transverse speed, the time since closest approach was at least 20 Myr 
ago, but the latter transverse speed would imply a time as long as 50 Myr.
Since the collision is presumed to be off center, the shocks in the ISM
of the ring may have taken up to $\sim$10 Myr to fully develop and
lead to copious star formation (M. Krumholz, private communication).
Thus, the major epoch of star formation in the ring of Arp 147 likely
occurred some 10 to 40 Myr ago.

We have utilized available multiwavelength images of Arp 147 in the
NIR, optical, and UV, as well as new X-ray observations to better understand
the star formation and collision history of this interacting pair of
galaxies.

The blue ring is luminous in the NIR, optical, UV, and X-ray bands,
although its appearance changes in important ways from band to band.
A spatially-integrated spectrum of the ring is shown in Fig.\,\ref{fig:flam}.  
The $\lambda L_\lambda$ luminosity spectrum is roughly flat over the entire 
0.16-24 $\mu$m wavelength range (i.e., constant to within a factor of $\sim$2).  
The NIR emission is likely a manifestation
of star formation via heating of embedded dust by stellar UV (see \S
\ref{sec:SFR}).  The brightest NIR emission from the ring comes from
near the original nucleus whose reddish color in the HST image
indicates an abundance of interstellar dust.  The ring is lit up in the
UV nearly everywhere -- except in the vicinity of the original nuclear
site where it may be extinguished by dust.  It is quite blue in the
optical -- this is consistent with the detection in the UV.  Finally,
the luminous X-ray sources are distributed mostly around the northern
and eastern parts of the ring which also contain most of the bright
blue knots of star formation.

\begin{figure}[t]
\begin{center}
\includegraphics[width=0.99 \columnwidth]{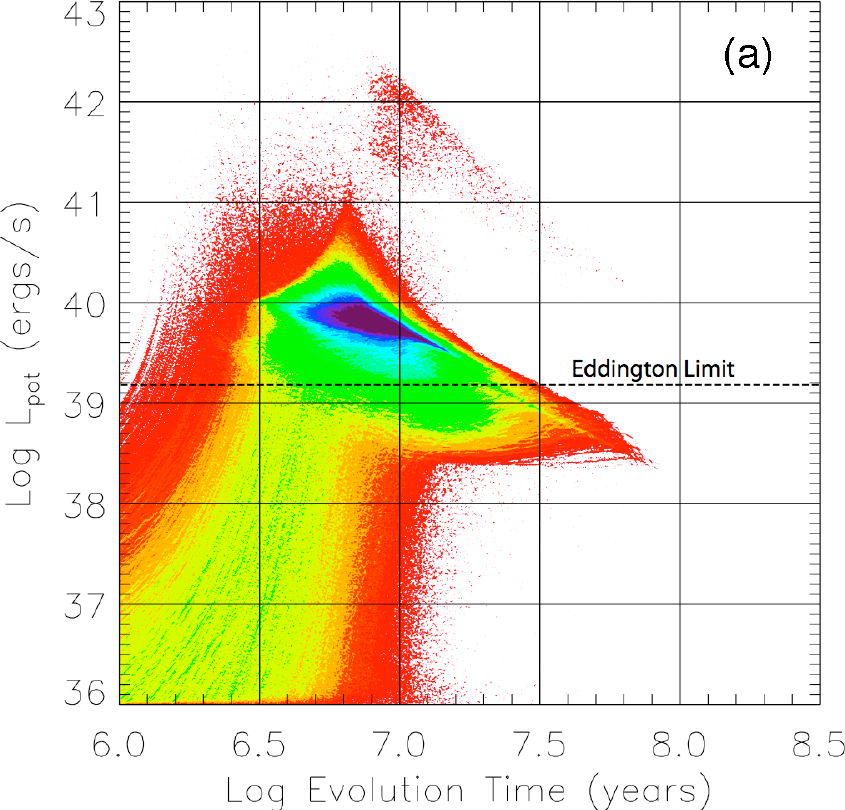}\vglue6pt 
\hglue3pt
\includegraphics[width=0.99 \columnwidth]{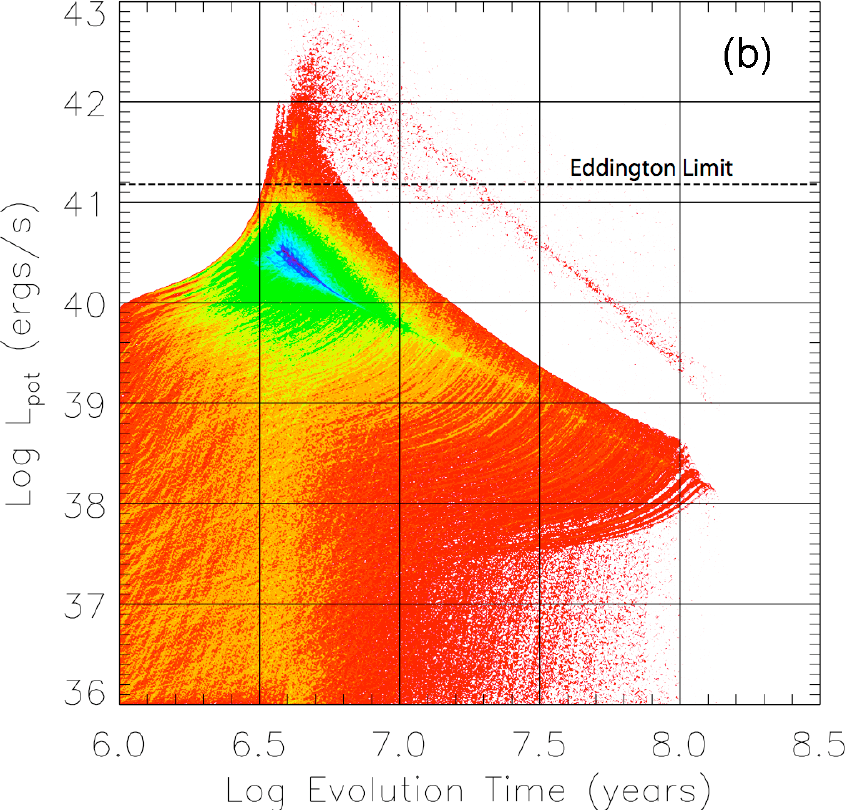}
\caption{Population synthesis of ULX binaries (adapted from Madhusudhan 
et al.\,2009).  Top panel: black-hole accretors with masses in the range
of 6-24 $M_\odot$ and donor stars of initial mass $10-34\,M_\odot$.
Bottom panel: intermediate mass black-hole accretors (1000 $M_\odot$)
donor stars of initial mass $5-50\,M_\odot$. The plot comprises the
evolution tracks of 30,000 binary systems; all assumed to be born
during the same starburst event.  ``Potential'' X-ray luminosity is
plotted against evolution time (time since the starburst).  The
potential luminosity is defined to be $L_{\rm pot} = \eta \dot M c^2$,
where $\dot M$ is the mass transfer rate, and $\eta$ is the efficiency
of the conversion of rest mass to energy of the accreted material.
During the evolution calculations, any mass that is transferred in
excess of the Eddington limit is ejected from the system.  The
potential luminosity is shown here to illustrate how high the
luminosity might reach in the absence of the Eddington limit.  In
panel (a) [panel (b)] the line marked ``Eddington Limit" is for a
$10\,M_\odot$ [$1000\,M_\odot$] black hole accreting H-rich
material. The color shading is related to the number of systems at a
given evolution time having a given luminosity -- blue being maximum
and red being minimum, in the ratio of $\sim$200:1. Very few X-ray
sources with $L_x \gtrsim 10^{40}$ ergs s$^{-1}$ ($L_x \gtrsim
10^{39}$ ergs s$^{-1}$) remain after $\sim$15 Myr (45 Myr).}
\label{fig:BPS}
\end{center}
\end{figure}

The intruder galaxy is quite bright in the optical and
NIR up to $\sim$ 5.8 $\mu$m, but then fades at the longer wavelengths
(see Fig.\,\ref{fig:flam}). The {\em Galex} image of the intruder shows little 
detectable UV radiation.  The integrated spectrum of the intruder is summarized
in Fig.\,\ref{fig:flam}.  If we utilize the expressions given in \S4 that relate the
spectral luminosities at 0.2, 8, and 24 $\mu$m to the star formation rate, we find
that the intruder has a SFR that is about an order of magnitude lower than that 
of the ring.  We interpret the bright NIR emission of the intruder as being due to 
dust and PAHs that are heated by ordinary star light (see, e.g., Draine \& Li 2007)
as well as to direct starlight.  There is also a luminous X-ray source which we associate
with the nucleus of the intruder galaxy.  Its luminosity of $1.4 \times
10^{40}$ ergs s$^{-1}$ is in the ULX range, but there is insufficient
angular resolution to tell whether this source is truly nuclear or
sufficiently off the nucleus to qualify as a ULX.  The exact dynamical 
center of the nucleus of the intruder is difficult to define to better
than $\sim$0.25$''$ ($\sim$150 pc) due to its irregular shape.

The X-ray luminosities of the 9 detected X-ray sources in the ring
galaxy are in the range $1.4-7 \times 10^{39}$ ergs s$^{-1}$.  Seven of
these sources qualify as ULXs, but none has a luminosity that exceeds
$10^{40}$ ergs s$^{-1}$.  In contrast, 9 sources of 21 detected in the
{\em Chandra} image of the main ring of the Cartwheel galaxy 
(Gao et al.~2003; Wolter et al.~2006) have $L_x$
in excess of $10^{40}$ ergs s$^{-1}$ even though the distance of the
Cartwheel is comparable to that of Arp 147 and the image of the
Cartwheel was obtained in a 75 ks exposure. Notwithstanding the small
number statistics of the luminous X-ray sources in the Arp 147 blue
ring, the absence of high-end ULXs in Arp 147 is significant
relative to the number seen in the Cartwheel.

The difference in the relative number of sources with $L_x \gtrsim
10^{40}$ ergs s$^{-1}$ suggests a difference in the star formation
histories of these galaxies.  In the Cartwheel it appears that a shock
wave is moving radially outward in the disk and star formation has
occurred over a long interval (i.e., hundreds of Myr) along or just behind 
the leading edge of the wave and is still ongoing at present.  By contrast, 
we argue below that the absence of extremely luminous X-ray sources in 
Arp 147 indicates that the peak of star formation therein occurred some
tens of Myr in the past and has declined sharply as of
$\sim$15 Myr ago.

ULXs are likely to evolve significantly on time scales of millions to
tens of millions of years.  This is demonstrated in
Fig.\,\ref{fig:BPS}, which shows a sample of the results of the population
synthesis study of candidate ULXs carried out by Madhusudhan et al.~(2008).
The candidate ULXs were (1) conventional high-mass X-ray binaries
consisting of a massive donor star initially of mass $10-34~M_\odot$
and an accreting stellar-mass, i.e., $6-24\,M_\odot$, black hole, and
(2) donor stars initially of mass $5-50~M_\odot$ feeding a
$1000~M_\odot$ black hole (IMBH).  All systems were assumed to be born
at the same time during a starburst event.  In Fig.\,\ref{fig:BPS} the
``potential'' X-ray luminosity (see the caption for the definition) is
plotted against evolution time (time since the starburst)\footnote{We
also indicate on Fig.\,\ref{fig:BPS} the location of the typical
Eddington luminosities.  In the case of the stellar-mass black hole
accretors we illustrate $L_{\rm Edd}$ for a $10 \, M_\odot$ black hole
accreting H-rich material.  A $20 \, M_\odot$ black hole accreting
He-rich material (later in the binary evolution) would have
$L_{\rm Edd}$ of $\sim$$5 \times 10^{39}$ ergs s$^{-1}$.}.  The color
shading is related to the number of systems at a given age (i.e.,
evolution time) having a given luminosity.  Some 30,000 binary
evolution tracks are included in each set of simulations.

The simulations of IMBH ULXs show X-ray luminosities that reach or
exceed $10^{40}$ ergs s$^{-1}$ for only about 15 Myr. Few ULXs remain
after $\sim$30 Myr. The simulations of accreting stellar mass black
holes yield similar conclusions assuming that either the Eddington
limit is not effective, or that the emission can be rather anisotropic, i.e., 
beamed. 

When compared with the maximum luminosities of the X-ray sources in
the ring of Arp 147, the simulation results suggest that the star
formation rate in the ring has not been high enough during the past $\sim$15
Myr to produce ULXs with $L_x \gtrsim 10^{40}$ ergs s$^{-1}$.
This is consistent with the inferred time since the collision.
This also suggests that the geometry is not like that of the
Cartwheel.  Rather than a disk with a propagating ring of star formation, 
Arp 147 is likely a small, tidally elongated and twisted galaxy that does 
not have such a radially propagating ring of star formation. 
Measurements of radial velocities at a number of positions around the ring 
would help to understand the true geometry.

\acknowledgments
We thank Philip Appleton and Mark Krumholz for helpful discussions.  We acknowledge {\em Chandra} grant GO0-11107X for partial support of this work.

\end{document}

%% file: Arp147.tbl.tex
\begin{deluxetable*}{ccccccccc}
\tablecolumns{8}
\tabletypesize{\scriptsize}
\tablecaption{Arp 147 X-Ray Source Properties\tablenotemark{a}}
\tablehead{
	\colhead{Source} &
	\colhead{Right Ascension} &
	\colhead{Declination}  & 
	\colhead{$\Delta \theta$\tablenotemark{b}} &
	\colhead{Net Counts} &
         \colhead{Power law} &
         \colhead{Med. Energy} &
	\colhead{$F_x$\tablenotemark{c}} &
	\colhead{$L_x$\tablenotemark{d}} \\
	\colhead{} & 
	\colhead{J2000 (h:m:s)} &
	\colhead{J2000 ($^\circ$:${'}$:${''}$)} & 
         \colhead{(\arcsec)}&
	\colhead{(0.5--8 keV)} &
	\colhead{photon index} &
	\colhead{keV} &
	\colhead{(0.5--8 keV)} & 
	\colhead{(0.5--8 keV)} 
}
\startdata
 Ring 1 & 03:11:18.04 & +01:19:02.3 & 0.10 & 14.7\err{4.2}{3.5} &2.0\err{0.5}{0.5} & $1.34 \pm 0.23$ & 2.9\err{1.3}{0.9} & 6.2\err{2.8}{1.9} \\
 Ring 2 & 03:11:18.34 & +01:19:02.4 & 0.15 &  6.7\err{3.0}{2.3} &2.2\err{0.7}{0.9} & $1.04 \pm 0.36 $ & 1.3\err{1.1}{0.6} & 2.7\err{2.4}{1.3} \\
 Ring 3 & 03:11:18.57 & +01:19:02.0 & 0.22 &  2.7\err{2.1}{1.4} &1.5\err{1.0}{1.0} & $1.82 \pm 0.60$ & 0.68\err{0.9}{0.4} & 1.4\err{1.8}{0.8} \\
 Ring 4 & 03:11:18.69 & +01:19:00.8 & 0.20 &  3.7\err{2.4}{1.7} &2.5\err{1.0}{1.1} & $1.25 \pm 0.50$ & 0.76\err{0.9}{0.4} & 1.6\err{1.9}{0.9} \\
 Ring 5 & 03:11:18.80 & +01:18:57.7 & 0.14 &  7.4\err{3.2}{2.5} &0.6\err{0.7}{0.6} & $1.01 \pm 0.36$ & 2.4\err{1.7}{1.0} & 5.1\err{3.6}{2.1} \\
 Ring 6 & 03:11:18.66 & +01:18:54.4 & 0.18 &  4.5\err{2.6}{1.9} &2.4\err{0.9}{1.0} & $1.35 \pm 0.48$ & 0.9\err{0.9}{0.5} & 1.9\err{2.0}{1.0} \\
 Ring 7 & 03:11:18.69 & +01:18:51.5 & 0.11 & 10.6\err{3.7}{3.0} &0.9\err{0.5}{0.5} & $1.70 \pm 0.29$ & 3.2\err{2}{1.2} & 6.9\err{4.2}{2.6} \\
 Ring 8 & 03:11:18.47 & +01:18:48.2 & 0.11 & 11.7\err{3.8}{3.1} &1.2\err{0.5}{0.5} & $1.41 \pm 0.28$ & 3.2\err{1.8}{1.2} & 6.7\err{3.9}{2.5} \\
 Ring 9 & 03:11:19.14 & +01:18:53.6 & 0.12 & 10.5\err{3.7}{3.0} &1.3\err{0.5}{0.5} & $1.75 \pm 0.29$ & 2.7\err{1.7}{1.0} & 5.7\err{3.6}{2.2} \\
 Galaxy & 03:11:19.51 & +01:18:48.4 & 0.06 & 41.7\err{6.8}{6.1} &2.9\err{0.3}{0.3} & $0.97 \pm 0.13$ & 6.8\err{1.2}{1.0} & 14\err{3}{2} \\
 Quasar & 03:11:20.05 & +01:18:58.9 & 0.05 & 68.7\err{8.6}{7.9} &1.4\err{0.2}{0.2} & $1.48 \pm 0.09$ & 17\err{3.0}{2.5} & NA \\ 
   Star & 03:11:20.41 & +01:18:41.7 & 0.11 & 12.6\err{4.0}{3.3} &2.1\err{0.5}{0.6} & $1.22 \pm 0.25$ & 2.5\err{1.2}{0.8} & NA 
\enddata
\tablenotetext{a}{All confidence intervals are 1$\sigma$.}
\tablenotetext{b}{$\Delta \theta$ is the relative uncertainty in the source location.}
\tablenotetext{c}{Units for the X-ray flux, $F_x$, are $10^{-15}$ ergs cm$^{-2}$ s$^{-1}$.}
\tablenotetext{d}{Units of the X-ray luminosity, $L_x$, are $10^{39}$ ergs s$^{-1}$.}
\label{tab:xray}
\end{deluxetable*}